\documentstyle[12pt]{article}
\renewcommand{\tiny}{\rm}
\newcommand{\be}{\begin{equation}}
\newcommand{\ee}{\end{equation}}
\newcommand{\bq}{\begin{eqnarray}}
\newcommand{\eq}{\end{eqnarray}}
\newcounter{saveeqn}
\newcounter{App} 

\begin{document}

\pagestyle{empty}
\hskip3.5truein{DFUPG 70/92}

\begin{center}
{\large \bf CHIRAL DYNAMICS  AND FERMION MASS GENERATION IN THREE
DIMENSIONAL GAUGE THEORY}\\
\vspace{1 cm}
{\large M. C. Diamantini}$^{\mbox{\tiny 1}}$ {\large and P.
Sodano}\footnote{This work is supported in part by a grant from the
M.U.R.S.T.}\\
\vspace{0.4 cm}
{Dipartimento di Fisica and Sezione I.N.F.N., Universit\'a di Perugia\\
Via A. Pascoli, 06100 Perugia, Italy}\\

\vspace{0.5 cm}
{\large G. W. Semenoff}\footnote{This work is
supported in part by the
Natural Sciences and Engineering Research Council of Canada.  G.S.
acknowledges the hospitality of the Physics Department of the
University of Perugia and I.N.F.N., Sezione di Perugia
where this work was completed.}\\
\vspace{0.4 cm}
{ Department of Physics, University of British Columbia\\
Vancouver, B.C., Canada V6T 1Z1}
\end{center}
\vskip 0.5truein
\begin{center}
{\bf Abstract}
\end{center}
We examine the possibility of fermion mass generation in 2+1--dimensional
gauge theory from the current algebra point of view.
In our approach the critical behavior is governed by the fluctuations of
pions which are the Goldstone bosons for chiral symmetry breaking.  Our
analysis supports the existence of an upper critical $N_F$ and exhibits the
explicit form of the gap equation as well as the form of the critical
exponent for the inverse correlation length of the order parameter.
\newpage
\pagestyle{plain}
\setcounter{page}{1}
The possibility of dynamical generation of fermion masses
is of fundamental importance to our present view
of quantum chromodynamics (QCD) as a theory of the strong interactions.
Central to our understanding of this phenomenon is the existence of a
critical coupling. When fermions have a sufficiently strong, attractive
interaction there is a pairing instability and the resulting condensate
breaks some of the flavor
symmetries, generates quark masses
and represents chiral symmetry in the Nambu-Goldstone mode.
This idea dates back to the earliest models for
chiral symmetry
breaking \cite{njl} and is prevalent in the
modern literature \cite{llm}.

Recently the issue of critical coupling
has been investigated in 2+1 dimensional gauge theories
\cite{anw,pisarski,pw}.  These theories provide
toy models which exhibit simpler behavior than
their 3+1--dimensional relatives \cite{jacktemp,appelquist}
and are also of interest
as effective field theories for some condensed matter systems
\cite{fradbk}.
Typically, their dimensionless expansion parameter
is $1/N_F$, where
$N_F$ is the number of quark flavors \cite{anw}.
Using Schwinger-Dyson equations in the $1/N_F$ approximation
for QED and QCD, \cite{anw} have found that there is a
critical $N^{\rm crit}_F$ such that
when $N_F<N_F^{\rm crit}$ chiral symmetry is
broken and when $N_F>N_F^{\rm crit}$ it is not broken and quarks remain
massless.

In the case of QED, this result has been the subject of some debate
\cite{pisarski,pw,j,ajm}.
Either an improved ladder approximation \cite{pw,ajm} or
renormalization group computation \cite{pisarski} find no critical behavior
and that chiral symmetry is broken for arbitrarily large $N_F$.
There are, however, numerical simulations \cite{kogut} of 2+1--
dimensional QED which find an $N^{\rm crit}_F$ remarkably close to that
obtained by \cite{anw}.

In this Letter we shall present further support for the existence of
$N^{\rm crit}_F$.  We shall advocate a picture which
is complementary to that of
critical attractive quark-quark interactions and in which
the critical behavior of the chiral symmetry breaking
phase transition is governed by the fluctuations of
the pions which are the Goldstone bosons for broken
continuous flavor symmetries.  We shall argue that an upper critical $N_F$
is natural since the order parameter is renormalized by the  $N_F^2/2$
pions with classical coupling constant $\sim 1/N_FN_C$ where $N_C$ is the
number of quark colors.
Their fluctuations are strong enough to destroy the ordered state when
$N_F=N_F^{\rm crit}\sim N_C$.

We are partially motivated by our recent study of strong coupling gauge
theory \cite{s} on the lattice.
We showed that, in the strong coupling limit, a
Hamiltonian lattice gauge theory with $N_C$ colors and $N_F/2$ lattice
flavors of staggered fermions (because of fermion doubling this corresponds
to $N_F$ continuum flavors of 2--component spinors in 2+1--dimensions) is
effectively a $U(N_F/2)$ quantum antiferromagnet, with
representations determined by $N_C$ and $N_F$.
We also identified chiral symmetry
breaking with the formation of either commensurate
$U(1)$ charge density waves or $SU(
N_F/2)$ spin density waves, i.e. N\'eel order.  We found that the cases
where $N_F/2$ is an odd or even integer are quite different.

When $N_F/2$
is odd the strong coupling limit necessarily breaks chiral symmetry, no
matter how large $N_F$ is and the condensate is a $U(1)$ charge density
wave. The ground state has a staggered structure with
the $SU(N_F/2)$ representations given by the Young tableau with $N_C$
columns and $N_F/4+1/2$ rows (and U(1) charge 1/2) on even
sites and $N_C$ columns and $N_F/4-1/2$ rows (and U(1) charge -1/2)
on odd sites.
(This was strictly true for $U(1)$ gauge theory, and likely for $U(N_C)$
and $SU(N_C)$ gauge theory where it could be proved only with some additional
assumptions about translation invariance.)

We found that when $N_F/2$ is an even integer, there is no $U(1)$ charge
density wave and the representation at each site was given by the Young
tableaux with $N_C$ columns and $N_F/4$ rows.

In either case of even or
odd $N_F/2$, there could be N\'eel order, which also
breaks chiral symmetry.  Quantum antiferromagnets with the kinds of
representations we considered have been analyzed in \cite{read} where they
found that, for small enough $N_F$, the ground state is ordered.  Also, in
the generic case, when $N_F$ is increased there is a phase transition with
$N_F^{\rm crit}\sim N_C$ to a disordered state. In this picture, the large
$N_C$ limit is the classical limit where the N\'eel ground state is
favorable and the small $N_C$ and large $N_F$ limit is where fluctuations
are large and disordered ground states are favored.

For example, the $SU(2)$ antiferromagnet with spin j corresponds to
4-flavor QCD with color group U(2j) and in particular to QED when j=1/2.
It has a N\'eel ordered ground state for any j, corresponding
to chiral symmetry breaking in the strong coupling limit of QCD.
However, an $SU(N_F/2)$ antiferromagnet with $N_F$ a large multiple of
4 and in a  representation of $SU(N_F/2)$ given by a Young tableau with
a single column of $N_F/4$ boxes corresponds to strong coupling QED with
$N_F$ flavors of fermions.
It is known that the ground state of this system
is disordered with several competing
flux and dimer phases \cite{fradbk,read}.
For some intermediate $N_F$ between 2 and $\infty$
the antiferromagnet has a phase transition where N\'eel order is lost.

It is tempting to conclude that this critical behavior of antiferromagnets
is related to the critical behavior of continuum gauge theory
found in \cite{anw} using Schwinger-Dyson equations.
In the following we wish to examine this
question further in the continuum by analyzing the dynamics of the
effective field theory for pions in the phase with broken chiral symmetry.

Our lattice results seem to imply that there is a subtle
difference between the cases where $N_F$ is an even or odd multiple of 2.
It is not clear whether this is a fundamental difference in the continuum
theory too, or merely an artifact of forcing gauge theory
to live on a lattice.
The numerical lattice simulations of QED in \cite{kogut} use staggered
Euclidean fermions and can therefore only study the case when $N_F$ is a
multiple of 4.
For compact continuum QED we argued in \cite{s}
that, if one uses Polyakov's idea \cite{polyakov}
of taking the Georgi-Glashow model with
spontaneous symmetry breaking $SO(3)\rightarrow U(1)$ to obtain a photon
with compact $U(1)$ gauge group, then it is known \cite{ni}
that the minimal number
of 2-component spinors the photons can couple to,
consistent with gauge and parity invariance,
is 4. Thus, if continuum QED is to be compact, $N_F$ is a multiple of 4.
This could also apply to $U(N_C)$ gauge theory because of the $U(1)$
subgroup.
However, the only such restriction on QCD with $SU(N_C)$ gauge group
constrains $N_F$ to be even.

We shall consider QCD with gauge group $SU(N_C)$ and $N_F$
(=2$\times$integer) flavors of massless 2-component quarks in Euclidean space
\bq
S=\int d^3x\left( \frac{1}{4e^2\Lambda}\sum_{a=1}^{N^2_C-1}
F_{\mu\nu}^aF_{\mu\nu}^a+\sum_{\alpha
=1}^{N_F}\bar\psi^\alpha \gamma^\mu(i\partial_\mu+A_\mu)\psi^\alpha\right)
\label{action}
\eq
where $e^2$ is the dimensionless coupling constant and $\Lambda$ is the
ultraviolet cutoff.
This action has $U(N_F)$ global flavor symmetry and also a $Z_2$ parity
symmetry under the replacement $(A_1,A_2,A_3)(x)\rightarrow(-A_1,A_2,A_3)(
x')$, $\psi(x)\rightarrow\gamma_1\psi(x')$, $\bar\psi(x)\rightarrow-\bar\psi
(x')\gamma_1$ with $x'=(-x_1,x_2,x_2)$. We shall assume that the
ultraviolet regularization preserves parity.

An order parameter for the $U(N_F)\times Z_2$ symmetry breaking is the
quark bilinear
\bq
M^{\alpha\beta}(x)=\bar\psi^\alpha(x)\psi^\beta(x)
\eq
$M$ is a Hermitean matrix and transforms under $U(N_F)$ as
$M\rightarrow gMg^{\dagger}$ and under parity as $M\rightarrow -M$. ( This
is in contrast with its counterpart in
3+1--dimensions, $\mu=\bar\psi_L\psi_R$ which is a complex matrix and
transforms under $SU_R(N_F)\times SU_L(N_F)$ as $\mu\rightarrow g\mu
h^{\dagger}$.  )
Flavor symmetry breaking is governed by the effective Landau-Ginsburg
action
\bq
S_{eff}=\int d^3x{\rm ~tr}\left( c_1\partial_\mu M\partial_\mu M +c_2M^2+
c_3M^4+\ldots\right)
\label{gl}
\eq
Note that, in the large $N_C$ limit, the coefficients $c_i$ in (\ref{gl})
are fermion loops with meson operator insertions which are naturally of order
$N_C$ \cite{thooft}.

We shall consder the symmetry breaking pattern $U(N_F)\times Z_2\rightarrow
U(n)\times U(N_F-n)$.
In this case, $M$ has a constant vacuum expectation value
\bq
M_0=<M>={\rm~const~}{\rm diag~}\left(1,1,1,\ldots,-1,-1,-1\right)
\eq
where there are $n$ 1's and $N_F$-$n$ -1's. We shall show that
the case where $n=N_F/2$, for which a residual parity symmetry
can be defined, is dynamically favorable.

We are interested in the dynamics of Goldstone bosons which are described
by a sigma model with target space
the Grassmannian
$$
\frac{U(N_F)\times Z_2}{U(n)\times
U(N_F-n)} ~~~.
$$
With the ans\"atz $M(x)= g(x)M_0g^{\dagger}(x)$
we obtain the sigma model
\bq
S_{\rm eff}=\int d^3x \frac{\Lambda N_C}{2f^2}{\rm ~tr~}
\left(\bigl[g\partial_\mu g^{\dagger},M_0\bigr]
\bigl[g\partial_\mu g^{\dagger},M_0\bigr]\right)
+S_{WZ}
\label{s}
\eq
where we have renamed the coefficient, which here plays the role of
coupling constant and we have extracted its natural order in $N_C$.
$S_{WZ}$ is a  Wess-Zumino term which must be added to the sigma model
action in order to break an unwanted discrete symmetry and to obtain a
sensible current algebra \cite{fr}.
(\ref{s}) is equivalent to the gauged principal chiral model
\bq
S_{\rm eff}=\int d^3x \frac{\Lambda N_C}{f^2}{\rm~tr~}\left(
(Dg)^{\dagger}\cdot(
Dg)+i\lambda(g^{\dagger}g-1)\right) +S_{CS}[V,W]
\label{gs}
\eq
where $D=\partial +i(V+W)$ and
$V$ and $W$ are Hermitian gauge fields which have components in the upper
left $n\times n$ block and in the lower right $(N_F-n)\times(N_F-n)$
block respectively.  In (\ref{gs}) the Chern-Simons action is
\bq
S_{\rm CS}[V,W]=i\frac{\theta}{4\pi}\int d^3x{\rm tr}\left(VdV-WdW+
\frac{2}{3}V^3-\frac{2}{3}W^3\right)
\label{wz}
\eq
and we have introduced an $N_F\times N_F$ Hermitean
Lagrange multiplier field $\lambda$ to enforce the
constraint $gg^{\dagger}=1$.
Eliminating $V$ and $W$ in (\ref{gs}) using their equations of motion
yields (\ref{s}), up to higher derivative terms which come from
approximating the Wess-Zumino term by the Chern-Simons action.
In \cite{fr} it was argued that,
the model described by (\ref{gs}) has features remarkably similar to the
commonly accepted features of the Skyrme model of 3+1-dimensional QCD, such
as solitons which have Fermi statistics and behave like baryons.  They also
argued that to obtain the correct current-current commutation relation, one
must set $\theta=N_C$.  To be general, we shall keep $\theta$ arbitrary in
the following analysis.

In order to study the quantum properties of this model, we first note that
the field $g$ appears quadratically and can be
integrated to get the effective theory
\bq
S_{\rm eff}=N_F{\rm ~TR~}\ln(-D^2+i\lambda)-\int d^3x \frac{i
\Lambda N_C}{f^2}{\rm~tr~}\lambda+S_{CS}[V,W]+S_{\rm ghost}
\label{sp}
\eq
where, to fix the gauge freedom and properly define the quantum problem we
have added the Faddeev-Popov ghost action,
$$
S_{\rm ghost}=\int d^3x{~\rm tr}
\left( \frac{1}{2\alpha}(\partial V)^2+\frac{1}{2
\beta}(\partial W)^2+ \partial c^{\dagger}(\partial+iV)c
+\partial d^{\dagger}(\partial+iW)d\right)
$$

As in the standard approach to sigma models \cite{polyakov,arafeeva},
the remaining integral over $\lambda$, $U$ and $V$
is done by saddle-point approximation.  It is
assumed that the gauge fields are zero at the saddle point.  The saddle
point value of $\lambda$ provides a mass for the chiral field $g$ in
(\ref{gs}). When this mass
is non-zero $g$ fluctuates about $g$=0 and the model is
disordered.  When the saddle point is at zero, we obtain the ordered phase.
Assuming a constant saddle point and putting $i\lambda=\mu^2$,
we obtain the gap equation
\bq
\frac{N_FN_C}{f^2}\Lambda=\frac{N_F^2\Lambda}{2\pi^2}\left(1-\frac{
\mu}{\Lambda}\arctan\frac{\Lambda}{\mu}\right)
\eq
A solution of this equation exists and the model is disordered if
\bq
N_F\geq 2\pi^2\frac{N_C}{f^2}
\label{gap}
\eq
with the equality giving the condition for criticality.  For fixed $f^2$
and $N_C$ we can interpret this as an equation for critical $N_F$.  When
$N_F$ exceeds $N_F^{\rm crit}=2\pi^2N_C/f^2$, the chiral symmetry breaking
condensate is unstable.

Note that, unlike the case of the more conventional O(N) non--linear sigma
model, the saddle point approximation in the present case is not controlled
by any small parameter such as 1/N.  This is because, even though the
effective coupling constant of the saddle-point method
is $\sim 1/N_F^2$, there remain of $\sim
N_F^2$ degrees of freedom.  The correct large $N_F$ expansion of
(\ref{gs}) would involve the
topological expansion where, like the large $N_C$ expansion of
QCD \cite{thooft},
it is necessary to sum all planar graphs to obtain the leading order.

In spite of this, we have two reasons to believe in
the validity of (\ref{gap}).  First of all, as we shall see in the
following,
corrections from the next order in the saddle point approximation are
indeed smaller than the leading order.  Secondly, alternative to the saddle
point analysis, we could do a weak coupling expansion.  If $N_F$ is large
enough, criticality in (\ref{gap}) is obtained in the weak coupling
region.

The computation of the gap equation to next order is somewhat sophisticated.
We must expand (\ref{sp}) to quadratic order in $V,W,\lambda$,
drop linear
terms \cite{jackiw}
and perform the functional integral in the Gaussian approximation.
\bq
S_{\rm eff}=N_F^2V\int^\Lambda
\frac{d^3k}{(2\pi)^3}\ln(k^2+\mu^2)-\frac{\Lambda
N_CN_F\mu^2}{f^2}V+N_F^2V\int^\Lambda\frac{d^3k}{(2\pi)^3}\frac{1}{k^2+
\mu^2}+
\nonumber \\ +
N_F\int{\rm tr}\left(\frac{1}{2}
\lambda\Delta\lambda+\frac{1}{4}F^V\Pi F^V+\frac{1}{4}F^W
\Pi F^W+\dots\right)+S_{CS}+S_{\rm ghost}
\eq
where
$$
\Delta(k,\mu)= \frac{1}{4\pi\vert k\vert}\arctan\frac{\vert k\vert}{2\vert\mu
\vert}~~~,
$$
$$
\Pi(k,\mu)=\frac{1}{8\pi}\left(\frac{2\vert\mu\vert}{k^2}+\frac{k^2+4
\mu^2}{\vert k\vert^3}\arctan\left(\frac{\vert k\vert}{2\vert\mu\vert}
\right)\right)~~~,
$$
$V$ is the volume, we have set $\lambda\rightarrow -i\mu^2-i\delta
\mu^2+
\lambda$ and $\delta\mu^2=-2\Lambda^2/\pi^2$ is a counterterm which arises
from an additive
shift of $\lambda$ and which is necessary to cancel a quadratic divergence.
This infinite shift of the Lagrange multiplier field
is a standard feature of sigma model renormalization \cite{arafeeva}.
The result of integrating the Gaussian fluctuations is
$$
S_{\rm eff}/V=-\frac{\mu^2N_FN_C}{f^2}\Lambda+
\int^\Lambda \frac{d^3k}{(2\pi)^3}\left(N_F^2\ln(k^2+\mu^2)+N_F^2\delta\mu^2
\frac{1}{(k^2+\mu^2)}+\right.
$$
$$
\left.~~~+
\frac{N_F^2}{2}\ln\Delta+\frac{n^2+(N_F-n)^2}{2}\ln\left(
k^2\Pi^2+(\theta/4\pi)^2\right)
+\ldots\right)
$$
\bq
=\left(-\frac{N_C}{N_Ff^2}+\frac{1}{2\pi^2}
-\frac{1}{\pi^4}+\frac{2}{\pi^2}\frac{1+2n^2/N_F^2-2n/N_F}{1+
(8\theta/\pi^2)^2}\right)N_F^2\Lambda\mu^2
-
\nonumber\\
-\frac{\pi}{3}\left(1
-\frac{2\pi^2-8}{\pi^4}\ln\frac{\Lambda}{\mu\alpha}
+\frac{16}{\pi^4}\frac{1+2n^2/N_F^2-2n/N_F}
{1+(8\theta/\pi)^2}\ln\frac{\Lambda}
{\mu\beta}\right)N_F^2\vert\mu\vert^3+\ldots
\label{effpot}
\eq
Here, $\alpha$ and $\beta$ are (unknown) constants.  The method for
computing the integrals in described in the Appendix.
If interpreted as an effective potential for $\mu$, $S_{\rm
eff}$ is upside--down and apparently unstable. This originates  with
subtleties in dealing with complex saddle points.  We refer the reader to
the standard literature on the subject \cite{bender}.
The corrected formula for the critical line is
$$
N_F^{\rm crit}=2\pi^2\frac{N_C}{f^2}\left(
1-\frac{2}{\pi^2}+\frac{4(1+2n^2/N_F^2-2n/N_F)}{1+(8\theta/\pi)^2}\right)^{-1}
$$
When $\theta$ (which is equal to $N_C$)
is large, the right hand side differs by about 20 percent
from the leading order estimate.  The largest $N_F^{\rm crit}$ occurs when
$n=N_F/2$.  Thus, the first and therefore most
stable ordering which
occurs as as we lower $N_F$ is the parity symmetric phase $n-N_F/2$.

We can also interpret the logarithms in (\ref{effpot})
as changing the exponent,
$
\vert\mu\vert^3\rightarrow\vert\mu\vert^{3(1+\gamma)}
$,
where
$$
\gamma=\frac{8}{3\pi^4}\left(1-\frac{\pi^2}{4}+\frac{1+2n^2/N_F^2-2n/N_F}
{1+(8\theta/\pi)^2}\right)
$$
and the gap equation has solution
$$
\mu\sim (N_F-N_F^{\rm crit})^{1/(2+3\gamma)}
$$
which exhibits the critical exponent (again a small correction of the
leading order result 1/2) for the inverse correlation length of the
order parameter.

In conclusion, we note that
the beta function of the Grassmannian sigma model has been computed using
the $\epsilon$--expansion about 2 spacetime dimensions in
\cite{brezin}.  In their notation it is
$$
\beta(t)=(d-2)t-N_Ft^2-...
$$
and has an ultraviolet stable fixed point at $t=(d-2)/N_F$.
In our notation the coupling constant is
$t=f^2/2\pi^2N_C$, and in the leading order, the critical point (\ref{gap})
occurs at the zero of the beta--function.
This is an ultraviolet stable fixed point (it cannot be infrared
stable since the flow to low momentum should be interrupted by mass
generation).

\noindent
{\bf Appendix: Calculation of Integrals}

As an example of the integrations needed to find the quantum corrections to
the gap equation, consider the integral
$
I[\Lambda/\mu]=\int_0^{\Lambda/2\mu} dx~x^2\ln\frac{2}{\pi}\arctan x
$
which we want to evaluate in the limit $\Lambda/\mu\rightarrow \infty$.  To
evaluate the integral, we first extract the divergent parts as
$$
I[\Lambda/\mu]=\int_0^{\Lambda/2\mu} dx~x^2\left(\ln\frac{2}{\pi}\arctan x
+\frac{2}{\pi x}+\frac{1}{\pi^2 x^2}-\frac{2\pi^2-8}{3\pi^3x^2(x+\alpha)}
\right)+
$$
$$
-\frac{\Lambda^2}{4\pi\mu^2}-\frac{\Lambda}{\pi^2\mu}+\frac {2\pi^2-8}
{3\pi^3}\ln(\Lambda/2\mu\alpha)
$$
In the first term the infinite cutoff limit can safely be taken, to get a
function of $\alpha$.  However, $\alpha$ is an arbitrary number, so the
integral can be parameterized as
$$
I[\Lambda/\mu]=-\frac{\Lambda^2}{4\pi\mu^2}
-\frac{2\Lambda}{\pi^2\mu}+\frac {2\pi^2-8}
{3\pi^3}\ln(\Lambda/2\mu\tilde\alpha)
$$
where $\tilde\alpha $ is a fixed, but unknown (and irrelevant) constant.

\end{document}